\documentclass[lettersize,journal]{IEEEtran}
\usepackage{amsmath,amsfonts}
\usepackage{physics,amsmath}
\usepackage{algorithm}
\usepackage{algpseudocode}
\usepackage{array}
\usepackage{bm}
\usepackage{subcaption}

\usepackage{mathtools}
\usepackage{xcolor}
\usepackage{subcaption}
\usepackage{textcomp}
\usepackage{stfloats}
\usepackage{lipsum}  
\usepackage{url}
\usepackage{verbatim}
\usepackage{graphicx}
\usepackage{cite}
\usepackage{pdfpages}
\usepackage{tikz}
\usetikzlibrary{3d}
\usetikzlibrary{arrows.meta} 

\DeclareCaptionLabelFormat{simple}{#2}
\begin{document}

\title{Ion Transmitter for Molecular Communication\\

\thanks{}
}

\author{Shaojie Zhang,~\IEEEmembership{Student             Member,~IEEE}
        and Ozgur B. Akan,~\IEEEmembership{Fellow,~IEEE}
        \thanks{The authors are with the Internet of Everything Group, Electrical
        Engineering Division, Department of Engineering, University of Cambridge,
        CB3 0FA Cambridge, U.K. (e-mail: sz466@cam.ac.uk, oba21@cam.ac.uk). }
        \thanks{Ozgur B. Akan is also with the Center for neXt-Generation
        Communications (CXC), Department of Electrical and Electronics Engineering, Ko\text{\c{c}} University, 34450 Istanbul, Turkey (e-mail: akan@ku.edu.tr)}
        \thanks{This work was supported in part by the AXA Research Fund (AXA Chair
for Internet of Everything at Ko\text{\c{c}} University).}
}
\maketitle

\begin{abstract}
Molecular communication (MC) is an emerging paradigm that takes inspiration from biological processes, enabling communication at the nanoscale and facilitating the development of the Internet of Bio-Nano Things (IoBNT). Traditional models of MC often rely on idealized assumptions that overlook practical challenges related to noise and signal behavior. This paper proposes and evaluates the first physical MC ion transmitter (ITX) using an ion exchange membrane. The circuit network model is used to simulate ion transport and analyze both transient and steady-state behavior. This analysis includes the effects of noise sources such as thermal and shot noise on signal integrity and SNR. The main contributions of this paper are to demonstrate how a practical MC ITX can produce a realistic waveform and to highlight future research challenges associated with a physical membrane-based ITX.
\end{abstract}


\section{Introduction}

\IEEEPARstart{M}{olecular} communication (MC) is inspired by nature, mimicking the way living cells encode, transmit, and receive information. Compared to other communication methods, MC exhibits better biocompatibility, higher energy efficiency, and greater robustness within physiological conditions. Therefore, it has become one of the most promising methods for enabling nanonetworks and the Internet of Bio-Nano Things (IoBNT) \cite{akyildiz2015internet}. Novel applications can be realized through MC, which has the potential to revolutionize our understanding of disease mechanisms by actively linking information and communication technology (ICT) with other biotechnologies, such as smart drug delivery, artificial organs, and lab-on-a-chip systems \cite{akan2017fundamentals}. As the Internet of Things (IoT) evolves, biological entities are expected to become integrated into its framework, with MC providing a cyber-physical interface that extends connectivity into new territories \cite{akan2023internet}.

MC has been broadly studied from various theoretical aspects, such as channel modeling, modulation \cite{pierobon2013capacity}, and detection techniques \cite{kilinc2013receiver}. In the Internet of Nano-Things, nanomachines are considered potential receivers. Bio-transceivers and bio-nanomachines have been proposed in \cite{Nakano2014bionanomachine} and \cite{Unluturk2015bacteriareciver}, respectively. While creating nanomachines solely from biocomponents offers advantages in terms of biocompatibility, there are drawbacks when it comes to accomplishing nanonetworks for a wide range of applications. The low computational capabilities of biocomponents significantly limit the implementation of complex communication protocols and algorithms \cite{Unluturk2015bacteriareciver}. Additionally, biocomponents can operate only in \textit{in vivo} applications. Moreover, biocomponents pose challenges in seamlessly connecting biological entities to electronic entities such as computers, complicating integration into the broader IoT \cite{Akyildiz2015internetofbionanothings}.

On the other hand, artificial MC components, such as artificial receivers, can operate \textit{in situ} and conduct continuous, label-free operations for both \textit{in vivo} and \textit{in vitro} applications \cite{kuscu2019transmitter}, \cite{Kuscu2016MCreceiverbiosensor}. An artificial MC receiver selectively detects targeted information ligand concentrations and translates them into more understandable signals. However, the discrete nature of information molecules and the highly nonlinear, time-varying channel properties make reliable practical MC devices challenging, hindering the practical implementation of MC \cite{kuscu2019transmitter}.

In the last few decades, theoretical models of MC transmitters (TX) have been developed in various ways. The most common approach models the TX as a spherical-like entity that releases molecules into the environment, often assuming an idealistic scenario with a spontaneous release of molecules. However, in most natural systems, MC occurs via ion channels distributed across the cell surface, with the opening and closing of these channels controlled by complex biochemical processes \cite{Noel2016ChannelImpulse,Farsad2016MolecularCommunicationSurvey,Arjmandi2016IonChannelModulator}. Recent efforts have focused on developing more realistic TX models that consider molecule production, harvesting, and the finite size of both the TX and the information molecules. Some models incorporate the harvesting of molecules from the surrounding environment, treating the TX as a sphere with gates that control the flow of molecules. This approach includes a biochemical reaction model for molecule absorption, adding another layer of realism to the design \cite{Ahmadzadeh2022MoleculeHarvestingTransmitter}.

Further refinements have introduced concepts like functionalized nanoparticles with switchable membranes for controlling the release and reloading of information molecules, thus improving TX efficiency \cite{Schafer2022ControlledSignaling}. Despite their complexity, some promising designs have emerged, including hydrophobic nanopores and porous membranes for better control of molecules, as well as hydrogel materials that swell or deswell in response to stimuli, enabling controlled molecule release. These developments mark significant progress toward practical MC systems \cite{kuscu2019transmitter}. Unfortunately, due to the high complexity of physical TX architectures at the microscale, no physically feasible TX has been reported in the literature that satisfies all the TX design requirements.

Ion exchange membrane have been studied for the use of desalination and energy production. The most widely used theoretical approach involves a model where ionic transport is governed by the Nernst-Planck and Poisson equations \cite{manzanares1993numerical}. Due to the non-linearity nature of the Nernst–Planck and Poisson equation, an exact analytical solution for such membrane system is almost impossible to obtain. The network simulation method \cite{moya2001stationary,moya1995ionic,moya1999application,moleon2009transient} represents the physicochemical process in terms of a graphical model—analogous to an electrical circuit diagram—which can then be analyzed using standard circuit simulation software. This approach utilizes well-developed and commercially available tools for circuit analysis. As a result, the dynamic behavior of the entire membrane system can be determined without the explicit need to solve the governing differential equations. 

In this paper, the first membrane-based ion transmitter (ITX) is proposed and evaluated using the circuit network simulation methods through the actual waveform and fundamental noise performance, and pointing out some practical challenges in fulfilling such system. Sections \ref{Principles of Ion Exchange Membranes and Electrodialysis} covers the fundamental principles of ion exchange membranes and electrodialysis, detailing their role in ion transport and MC. The network simulation model, described in Section \ref{Network Simulation Model}, utilizes the Nernst-Planck and Poisson equations to represent ion transport. Section \ref{Model simulation} gives the simulation parameters and models key noise sources, including thermal and shot noise, and their effects on signal behavior. Section \ref{Results} presents model simulations showing ion flux behavior under different signal voltages, examining steady-state and transient responses. Finally, Section \ref{Future Direction and Conclusion} concludes with future research recommendations.

\section{Principles of Ion Exchange Membranes and ITX Design}
\label{Principles of Ion Exchange Membranes and Electrodialysis}
\begin{figure*}[t]
    \centering
    \includegraphics[width = 1\textwidth]{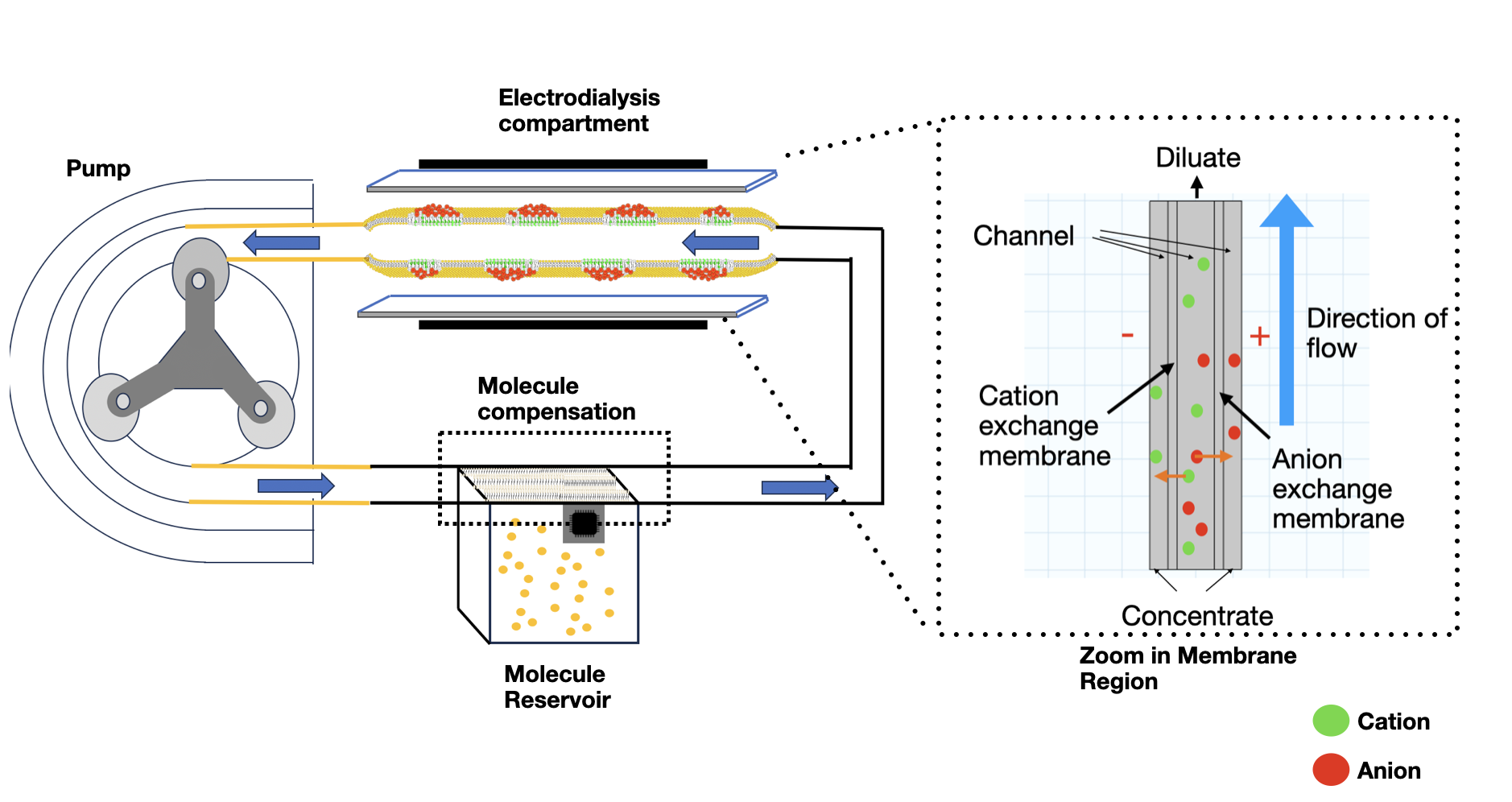}
    \caption{Proposed physical MC ITX.}
    \label{Design of physical transmitter}
\end{figure*}
Ion exchange membranes (IEMs) are barrier membranes developed to allow only the passage of ions of a specific charge. These include cation exchange membranes (CEMs) that allow positively charged ions (cations) to pass through as well as anion exchange membranes (AEMs) that transport negatively charged ions (anions). Various types of ion exchange membranes are reported, which obtain their selectivity owing to fixed charged groups embedded within their structure that repel ions of like charge and permit transport of species with the opposite charge \cite{xu2005ion}.

In electrodialysis, for example, a solution to the inorganic ion removal problem is achieved by applying an electric field across these membranes, causing ion migration. In this process, cations migrate to the negative electrode through cation exchange membranes, while anions move to the positive electrode through anion exchange membranes. The ability of this selective ion transport to separate and extract specific charged species from a solution has made electrodialysis essential for applications like water desalination, ion concentration, and purification \cite{ran2017ion}. From the MC perspective, ion transport can generate ion flux, which naturally aligns with the concentration shift keying (CSK) modulation scheme. For analysis purposes, we consider only half of the system due to its symmetry, treating cations as the positive ions that function as information molecules passing through the cation exchange membrane. 

Here we propose a new device based on such ion-exchange membrane. The design comprises three main components: a pump, which provides flow in the system to prevent fouling \cite{Campione2018ElectrodialysisWaterDesalination}; a drug reservoir and a electrodialysis compartment for information ions release, using an ion-exchange membrane, as shown in Fig. \ref{Design of physical transmitter}.

For on-off keying ion release control, maintaining a relatively stable ion concentration within the electrodialysis compartment is essential to ensure consistent output levels for the same driving voltage. The current passing through the electrodialysis compartment serves as a critical indicator of the concentration of information molecules, assuming these molecules exist in ionic form. In this study, we assume that the molecule reservoir is capable of automatically maintaining a steady concentration in the circulation channel, providing a stable baseline for reliable ITX operation. The two outer channels in the electrodialysis compartment are connected to the desired communication channel and serve as the ITX output.

\section{Network Simulation Model}
\label{Network Simulation Model}
\begin{figure*}[t]
    \centering
    \includegraphics[width = 1\textwidth]{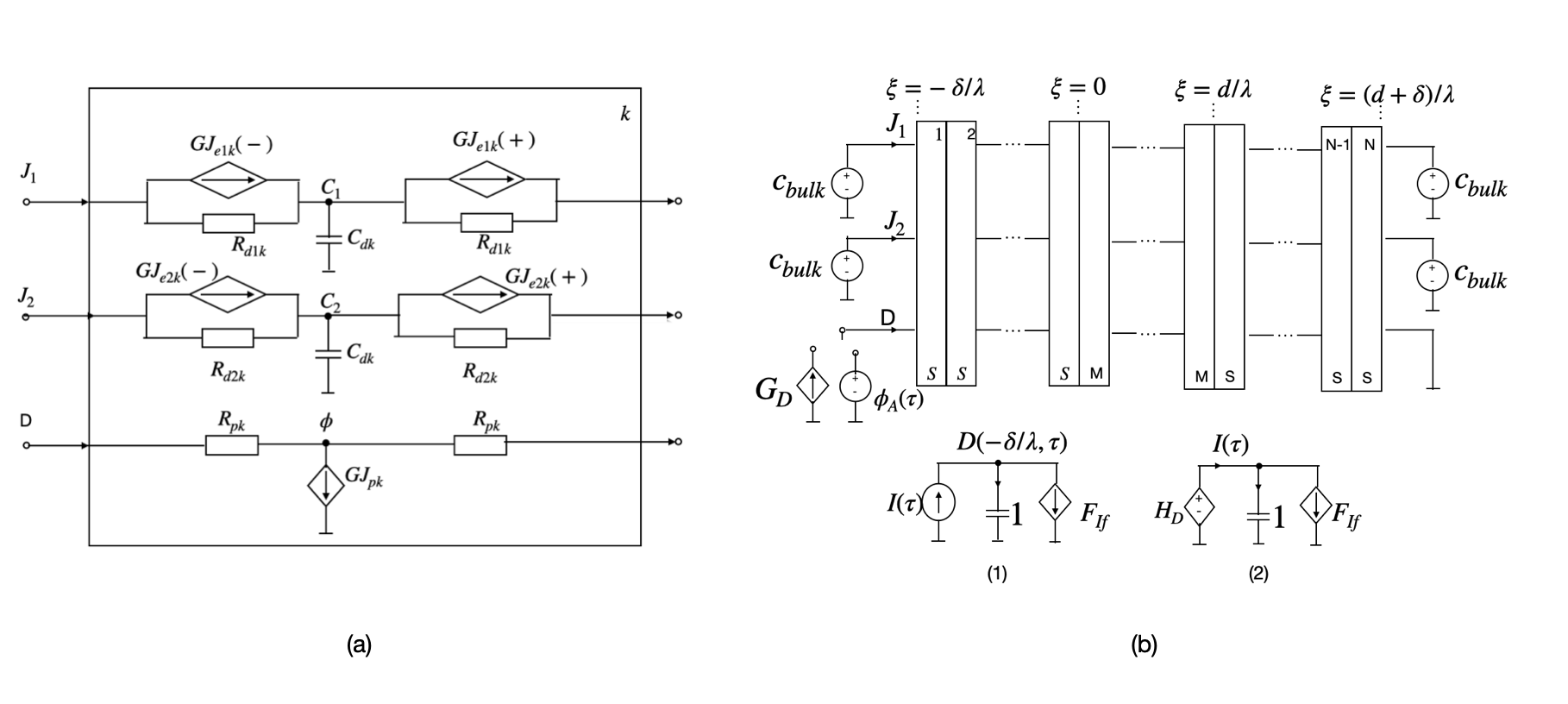}
    \caption{(a) Network model for the electrodiffusion in a volume element. (b) Network model for an ion-exchange membrane system. The details in boxes 1-N are provided in (a). Letters $S$ and $M$ indicate the solution and membrane regions, respectively.
    }
    \label{(a) Network model for the electrodiffusion in a volume element. (b) Network model for an ion-exchange membrane system. The details in boxes 1-N are provided in (a). Letters S and M indicate the solution and membrane regions, respectively}
\end{figure*} 
\subsection{Problem Formulation}
We consider a model system of an ion-exchange membrane that extends from \(x = 0\) to \(x = d\). The membrane is bathed by two bulk solutions, both with a concentration of \(c_{Bulk}\) and a layer width of \(\delta\) each. For a cation exchange membrane, we assume the membrane contains fixed negatively charged groups with a uniform concentration. We also assume that ion transport is one-dimensional and perpendicular to the membrane in the x-direction. In a membrane system, the transport of ions is governed by the Nernst-Planck equations, and the Poisson equation is used to describe the relationship between ion concentration and the electric field. For the mathematical convenience, these equations are given in the dimensionless form as \cite{moya1995ionic}
\begin{equation}
    \frac{\partial J_i(\xi, \tau)}{\partial \xi}=-\frac{\partial c_i(\xi, \tau)}{\partial \tau}, \quad i=1,2, ..., m ,
\label{conti}
\end{equation}
\begin{equation}
    J_i(\xi, \tau)=-D_{i p}\left[\frac{\partial c_i(\xi, \tau)}{\partial \xi}+z_i c_i(\xi, \tau) \frac{\partial \phi(\xi, \tau)}{\partial \xi}\right],
\end{equation}
\begin{equation}
    \frac{\partial \mathbf{D}(\xi, \tau)}{\partial \xi}=z_1 c_1(\xi, \tau)+z_2 c_2(\xi, \tau)-\theta(\xi)=\rho(\xi, \tau),
\end{equation}
\begin{equation}
    I(\tau)=z_1 J_1\left(-\frac{\delta}{\lambda}, \tau\right)+z_2 J_2\left(-\frac{\delta}{\lambda}, \tau\right)+\frac{\mathrm{d} \mathbf{D}\left(-\frac{\delta}{\lambda}, \tau\right)}{\mathrm{d} \tau},
    \label{eq2}
\end{equation}
where \( J_i(\xi, \tau) \) denotes the flux of ion species \( i \), \( c_i(\xi, \tau) \) represents the concentration of ion species \( i \). \( \phi(\xi, \tau) \) is the electric potential, and \( \rho(\xi, \tau) \) is the charge density. The parameters \( D_{ip} \) and \( z_i \) represent the diffusion coefficient and charge number of ion species \( i \), respectively, while \( \theta(\xi) \) is a background charge density function which can be expressed as  
\begin{equation}
    \theta(\xi)= \begin{cases}X, & 0 \leq \xi \leq d / \lambda \\ 0, & -\delta / \lambda<\xi<0 \text { and } d / \lambda<\xi<(d+\delta) / \lambda\end{cases},
\end{equation}
where $X$ is the dimensionless fixed negatively charged group concentration.

These quantities were normalized by the following relations:
\begin{equation}
    \begin{gathered}
\xi=\frac{x}{\lambda^{\prime}} ; \quad \tau=\frac{D_{\mathrm{a}} t}{\lambda^2} \\
c_i=\frac{c_i^{\prime}}{c_{\mathrm{a}}} ; \quad J_i=\frac{\lambda J_i^{\prime}}{D_{\mathrm{a}} c_{\mathrm{bulk}}} ; \quad D_{\mathrm{ip}}=\frac{D_{\mathrm{ip}}^{\prime}}{D_{\mathrm{a}}}, \quad i=1,2, ..., m;  \\
\phi=\frac{F \phi^{\prime}}{R T^{\prime}} ; \quad \epsilon=\frac{R T \epsilon^{\prime}}{F^2 c_{\mathrm{Bulk}} \lambda^2} ; \quad \mathbf{D}=\frac{\mathbf{D}^{\prime}}{F c_{\mathrm{Bulk}} \lambda^{\prime}} ;\\ \quad \rho=\frac{\rho^{\prime}}{F c_{\mathrm{Bulk}}} ; \quad X=\frac{X^{\prime}}{c_{\mathrm{Bulk}}};I=\frac{\lambda I^{\prime}}{F D_{\mathrm{a}} c_{\mathrm{Bulk}}},
\end{gathered}
\end{equation}
where the primed symbol are for the corresponding non-normalized variables. $D_{\mathrm{a}}, c_{\mathrm{Bulk}}$, and $\lambda$ are scaling factors with the dimensions of diffusion coefficient, molar concentration, and length, respectively. $D_{\mathrm{a}}, c_{\mathrm{Bulk}}$, and $\lambda$ are taken as characteristic values of the system studied. In particular, we have taken $\epsilon=1$, and so $\lambda$, given by $\lambda=\sqrt{\frac{\epsilon^{\prime} R T}{F^2 c_{\mathrm{Bulk}}}}$, which is the Debye length in the system. The constants $F$, $R$, and $T$ have their usual meanings.

For simplicity, there are only two types of ions exist in the membrane system, the valency are 1 and -1, respectively. The boundary conditions can be expressed as
\begin{equation}
    c_1\left(-\frac{\delta}{\lambda}, \tau\right)=c_1\left(\frac{d+\delta}{\lambda}, \tau\right)=c_{0},
\end{equation}
\begin{equation}
    c_2\left(-\frac{\delta}{\lambda}, \tau\right)=c_2\left(\frac{d+\delta}{\lambda}, \tau\right)=c_{0},
\end{equation}
\begin{equation}
    \frac{d \mathbf{D}\left(-\frac{\delta}{\lambda}, \tau\right)}{d \tau}=I(\tau)-z_1 J_1\left(-\frac{\delta}{\lambda}, \tau\right)-z_2 J_2\left(-\frac{\delta}{\lambda}, \tau\right)
    \label{eq1}
\end{equation}
\begin{equation}
    \phi\left(-\frac{\delta}{\lambda}, \tau\right)=\phi_{\mathrm{A}}(\tau),
\end{equation}
\begin{equation}
    \phi\left(\frac{d+\delta}{\lambda}, \tau\right)=0,
\end{equation}
where $c_0=c^{\prime}_0 / c_{bulk}$ the dimensionless concentration of the bulk electrolyte solution. The first two boundary conditions establish electrical neutrality at both boundaries. The third boundary condition, which follows directly from (\ref{eq2}), is the boundary condition on the time evolution of the electric displacement, $D$, at $n = -d/k$ and is applied when the total electric current density, $I(s)$, is the externally controlled variable. The rest conditions specify that a biased voltage has been applied to the system by setting a fixed value for the electric potential at the left boundary and grounding it at the right boundary.

\subsection{Network Model}
The network model represents the transport process that is obtained by dividing the physical region of interest, which we consider to have a unit cross-sectional area, into $N$ volume elements or compartments of width $\delta_k$ ($k = 1, \ldots, N$), small enough for the spatial variations of the parameters within each compartment to be negligible. 

The network model for the electro-diffusion process in a compartment is illustrated in Fig. 1(a), with a detailed explanation provided in elsewhere \cite{moya1995ionic}. The relation between those network elements and the parameters of the system can be obtained by

\begin{equation}
    R_{\mathrm{di} k}=\frac{\delta_k}{2 D_{i \mathrm{p}}},
\end{equation}

\begin{equation*}
J_{\text{eik}}( \pm)= \pm D_{i p} z_i c_i\left(\xi_k \pm \frac{\delta_k}{2}\right) \frac{\phi\left(\xi_k\right)-\phi\left(\xi_k \pm \frac{\delta_k}{2}\right)}{\delta_k / 2}, 
\end{equation*}
\begin{equation}
\text{where}
\begin{cases}
D_{i \mathrm{p}} = D_{i \mathrm{S}}, & \text{for the solution (S)} \\
D_{i \mathrm{p}} = D_{i \mathrm{M}}, & \text{for the membrane (M)}
\end{cases}
\end{equation}

\begin{equation}
    C_{\mathrm{d} k}=\delta_k,
\end{equation}
\begin{equation}
    R_{\mathrm{p} k}=\frac{\delta_k}{2 \epsilon},
\end{equation}
\begin{equation}
    G J_{\mathrm{p} k}=-\delta_k\left[z_1 c_1\left(\xi_k\right)+z_2 c_2\left(\xi_k\right)-\theta\left(\xi_k\right)\right],
\end{equation}
where $R_{\mathrm{d} i k}$ represents the resistor of the ion $i$'s diffusion inside compartment $k$, $G J_{\text {eik }}(\pm)$ is the voltage-controlled current source modeling the electrical contribution to the ionic flux, with the minus and plus signs indicating the flux entering and leaving the compartment $k$, respectively. $C_{\mathrm{d} k}$ is the capacitor representing the non-stationary effects of the electrodiffusion process in compartment $k$, $R_{\mathrm{p} k}$ is the resistor modeling the constitutive equation of the medium, and $G J_{\mathrm{p} k}$ is the voltage-controlled current source modeling the electric charge stored in compartment $k$.  

The time-independent concentrations of species with charges $z_1 = 1$ and $z_2 = -1$ are represented by independent voltage sources with values $c_{bulk}$ in the network model of Fig. \ref{(a) Network model for the electrodiffusion in a volume element. (b) Network model for an ion-exchange membrane system. The details in boxes 1-N are provided in (a). Letters S and M indicate the solution and membrane regions, respectively}. The electric potential origin is defined by short-circuiting the node $\phi$ at a specific point, $\xi = (d + \delta) / \lambda$. Subcircuit-1 models the boundary condition given by (\ref{eq1}) when switch-1 is activated, where an independent current source represents the electrical perturbation. The faradaic current is modeled using the current-controlled current source $F_{I f}$, while $C_1$ and $G_D$ represent the displacement current density through a capacitor and a voltage-controlled current source, respectively. The relationship between these network elements and the system parameters is provided as
\begin{equation}
    F_{l f}=z_1 J_1\left(-\frac{\delta}{\lambda}, \tau\right)+z_2 J_2\left(-\frac{\delta}{\lambda}, \tau\right),
\end{equation}
\begin{equation}
    G_D=\boldsymbol{D}\left(-\frac{\delta}{\lambda}, \tau\right).
\end{equation}
The voltage on $C_1$ is $\boldsymbol{D}\left(-\frac{\delta}{\lambda}, \tau\right)$, and $G_D$ takes this as the boundary condition for the Poisson equation. When switch-2 is activated, Subcircuit-2 computes the electric current density, assuming that the electric potential is the external perturbation. In this case, the perturbation comes from an independent voltage source $\phi_A(\tau)$. The faradaic current is modeled by $F_{\text{If}}$, and the displacement current density is represented by $C_1$ and $H_D$. The relationship between $H_D$ and system parameters is expressed as 
\begin{equation}
    H_D = \boldsymbol{D}\left(-\frac{\delta}{\lambda}, \tau\right),
\end{equation}
with the current through $H_D$ representing the total current density \cite{horno1996simulation}.

\subsection{Noise Source}
Here, we will discuss how noise could be part of the ITX system. The source of noise includes thermal noise and shot noise. The capacitance, $C_M$ $\text { (in units of } F^2 c_{\mathrm{bulk}} \lambda / R T \text { )}$, and resistance of the membrane, $R_M$ (in units of $\lambda R T /$ $F^2 D_{\mathrm{a}} c_{\mathrm{bulk}}$), within the network model can be expressed as
\begin{equation}
    C_{\mathrm{M}}=\frac{\epsilon \lambda}{d},
\end{equation}
\begin{equation}
    R_{\mathrm{M}}=\frac{d / \lambda}{z_1^2 D_{1 \mathrm{M}} c_1^*+z_2^2 D_{2 \mathrm{M}} c_2^*},
\end{equation}
where $c_1^*$, $c_2^*$ are equilibrium values of the cation concentration and the anion concentration \cite{moleon2009transient}.

Thermal noise, also referred to as Johnson-Nyquist noise, represents random ionic or electronic movements in a conductive medium that generate frequency-independent voltage fluctuations defined by a spectral density 
\begin{equation}
    S_{thermal}(\omega) = 4 k_B T \, \text{Re}(Z)^{\prime}\Delta f^{\prime},
\end{equation}
where \( \text{Re}(Z) \) is the dimensional real impedance and $\Delta f$ is the dimensional bandwidth. In the membrane's scaled network model, this adapts to 
\begin{equation}
    S_{thermal}(\omega) = \frac{4 k_B R_M }{\lambda R T c_{bulk} \Delta f } .
\end{equation}

Shot noise in membrane systems arises from the random fluctuations in ionic currents due to the discrete, quantal passage of ions through channels, causing variability in the number of ions crossing the membrane at any instant \cite{verveen1974membrane}. If we assume the total number of pores, $N$, and $k$ is the average number of ions passing through per second per pore, then the average number of ions passing through the membrane per second will be the average frequency of the Poisson wave, and in this case, it will also be the flux through the network model and can be given as
\begin{equation}
    J = N k.
\end{equation}

For a parallel RC membrane with time constant $\theta=R C$, the voltage response to the unitary impulse can be obtained as 

\begin{equation}
    g(\tau)= V_{0} e^{-\tau / \theta}, \tau \geq 0,
\end{equation}
where $V_{0}$ is the initial voltage amplitude across the membrane.
A poisson waves with exponential events can then be arrived at a fluctuating voltage, noise power spectral on voltage, $S_{shot}(\omega)$, as \cite{verveen1974membrane}
\begin{equation}
    S_{shot}(\omega)=\frac{J}{2 \pi } V_{0} \frac{\theta^2}{1+\omega^2 \theta^2}+(J V_{0} \epsilon R)^2 \delta(\omega).
\end{equation}

The voltage noise PSD could be transfer into flux noise PSD as follow. Assume small perturbations around the steady-state, such that
\begin{equation}
  \phi(\xi, \tau) = \phi_0 + \delta \phi(\xi, \tau),
\end{equation}
where \( \phi_0 \) is the steady-state values, \( \delta \phi \) is the fluctuations.

Substitute these into the flux equation and linearize to
\begin{equation}
    \delta J_i(\xi, \tau) \approx -D_{i \mathrm{p}} \left[z_i c_{i0} \frac{\partial \delta \phi(\xi, \tau)}{\partial \xi}\right].
\end{equation}

To simplify further, approximate the derivative \( \frac{\partial \delta \phi(\xi, \tau)}{\partial \xi} \) as \( \frac{\delta \phi(\xi, \tau)}{d} \), which yields:
\begin{equation}
    \delta J_i(\xi, \tau) \approx -D_{i \mathrm{M}} z_i c_{i0} \frac{\delta \phi(\xi, \tau)}{d}.
\end{equation}

The power spectral density \( S_J(\omega) \) for flux is related to the PSD \( S_\phi(\omega) \) of the voltage by:
\begin{equation}
    S_J(\omega)=\left(D_{i \mathrm{M}} z_i c_{i 0} \frac{1}{d}\right)^2 S_\phi(\omega) .
\end{equation}
The SNR for ITX will be
\begin{equation}
    SNR_{\text {out }}= J^2 / S_J(\omega).
\end{equation}

\section{Model simulation}
\label{Model simulation}
The network model is used to study the communication capability of homogeneous ion-exchange membrane systems by implementing it via an appropriate electric circuit simulation program. PSpice is used for this purpose. The system parameters we used are similar to those in \cite{moya1999application} to ensure good accuracy and CPU times. The spatial grid accounts for the membrane-solution interface, which spans about 4$\lambda$ in each phase. The other parameters follow \cite{manzanares1993numerical}, except for the charge number and diffusion coefficient of the mobile ions. Specifically, we used $X=1$, $z_1=1$, $z_2=-1$, $d=50 \lambda$, $\delta=100 \lambda$, $D_{1 \mathrm{~S}}=D_{2 \mathrm{~S}}=1$, $D_{1 \mathrm{M}}=D_{2 \mathrm{M}}=0.1$, $c^0=1$, $\epsilon=1$, $N=480$, and compartment thickness $\delta_k$ given by:
\begin{gather}
\delta_k = 3, \quad k = 1, \ldots, 30 \text{ and } 451, \ldots, 480, \\
\delta_k = 0.6,\\  \quad k = 31, \ldots, 40; 201, \ldots, 210; 271, \ldots, 280; 441, \ldots, 450, \\
\delta_k = 0.05, \quad k = 41, \ldots, 200 \text{ and } 281, \ldots, 440, \\
\delta_k = 1.5, \quad k = 211, \ldots, 270 .
\end{gather}
The initial conditions for the ionic concentrations and electric potential correspond to the equilibrium values, which must be previously calculated from a steady-state analysis for $I = 0$ \cite{moya2001stationary}. To apply further perturbation to the membrane system, external electric potential applied can be expressed as
\begin{equation}
    \phi_A(\tau)= \begin{cases}0, & \tau=0 \\ V_{sig}, & \tau>0\end{cases}.
\end{equation}

To derive the time constant for flux in a membrane system under the influence of an electric field, we start from the non-dimensionalized flux equation:

\begin{align}
J_i(\xi, \tau) = -D_{iM} \Bigg[ &\frac{\partial c_i(\xi, \tau)}{\partial \xi} + z_i c_i(\xi, \tau) \notag \\
&\times \frac{\partial \phi(\xi, \tau)}{\partial \xi} \Bigg], \quad i=1,2.
\end{align}

Substituting this expression into (\ref{conti}):
\begin{equation}
\frac{\partial c_i(\xi, \tau)}{\partial \tau} = D_{iM} \frac{\partial}{\partial \xi} \left[\frac{\partial c_i(\xi, \tau)}{\partial \xi} + z_i c_i(\xi, \tau) \frac{\partial \phi(\xi, \tau)}{\partial \xi}\right].
\end{equation}

Expanding the mixed derivative term using the product rule yields:
\begin{equation}
\begin{aligned}
\frac{\partial c_i(\xi, \tau)}{\partial \tau} = D_{iM} \Bigg[ & \frac{\partial^2 c_i(\xi, \tau)}{\partial \xi^2} + z_i \frac{\partial c_i(\xi, \tau)}{\partial \xi} \frac{\partial \phi(\xi, \tau)}{\partial \xi} \\
& + z_i c_i(\xi, \tau) \frac{\partial^2 \phi(\xi, \tau)}{\partial \xi^2} \Bigg].
\end{aligned}
\end{equation}

Assume small perturbations around a steady state:

\begin{equation}
c_i(\xi, \tau) = c_{i,0} + \delta c_i(\xi, \tau), \quad \phi(\xi, \tau) = \phi_0 + \delta \phi(\xi, \tau),
\end{equation}
where \( \delta c_i(\xi, \tau) \) and \( \delta \phi(\xi, \tau) \) are small perturbations. Substituting these into the expanded equation and retaining only first-order terms, we have:

\begin{equation}
\begin{aligned}
\frac{\partial \delta c_i(\xi, \tau)}{\partial \tau} = D_{iM} \Bigg[ & \frac{\partial^2 \delta c_i(\xi, \tau)}{\partial \xi^2} \\
& + z_i \left(\frac{\partial \delta c_i(\xi, \tau)}{\partial \xi} \frac{\partial \phi_0}{\partial \xi} \right. \\
& \left. + c_{i,0} \frac{\partial^2 \delta \phi(\xi, \tau)}{\partial \xi^2}\right) \Bigg].
\end{aligned}
\end{equation}

Assume that \( \frac{\partial \phi_0}{\partial \xi} \) is small or approximately constant over short distances, allowing \( \frac{\partial \delta c_i}{\partial \xi} \frac{\partial \phi_0}{\partial \xi} \) to be treated as a first-order perturbation. Additionally, assume \( c_{i,0} \) is uniform, so \( \frac{\partial c_{i,0}}{\partial \xi} = 0 \).

Under these assumptions, the PDE simplifies to:

\begin{equation}
\frac{\partial \delta c_i(\xi, \tau)}{\partial \tau} \approx D_{iM} \left[\frac{\partial^2 \delta c_i(\xi, \tau)}{\partial \xi^2} + z_i c_{i,0} \frac{\partial^2 \delta \phi(\xi, \tau)}{\partial \xi^2}\right].
\end{equation}

Focusing on the homogeneous part of the PDE for \( \delta c_i(\xi, \tau) \) under diffusion-only conditions (neglecting small \( \delta \phi(\xi, \tau) \) contributions), we obtain:

\begin{equation}
\frac{\partial \delta c_i(\xi, \tau)}{\partial \tau} \approx D_{iM} \frac{\partial^2 \delta c_i(\xi, \tau)}{\partial \xi^2}.
\end{equation}

The characteristic time constant for diffusion is:

\begin{equation}
\tau_{\text{flux}} \sim \frac{L^2}{D_{ip}},
\label{characteristic time constant}
\end{equation}
where \( L \) is the characteristic length scale over which \( \delta c_i(\xi, \tau) \) changes. This result shows that, under small perturbations, the original time constant for flux remains \( \tau_{\text{flux}} \sim \frac{L^2}{D_{iM}} \), even with the inclusion of non-uniform electric potential contributions as long as they are small.

\section{Performance Evaluation}
Key metrics like ion flux, concentration profiles, delay, spectral noise density, and signal-to-noise ratio (SNR) are used to assess the MC ITX's performance. The ion flux captures both transient and steady-state characteristics under different input signal conditions, offering information on the rate of molecule transport across the membrane. The spatial distribution of information molecules throughout the membrane system, which represents the ITX's capacity to produce distinct signaling patterns, is evaluated using concentration profiles. Delay, which indicates the time required for the TX to reach a steady state following a disturbance, is often overlooked in MC TX analyses. In this paper, the delay in the waveform response is numerically demonstrated. The noise power across various frequency ranges is provided by the analysis of spectral noise density, which quantifies the effect of thermal and shot noise on the signal integrity. All analyses are conducted within a dimensionless framework to align with the dimensionless network model, ensuring consistency and simplifying the interpretation of results.
\label{Results}
\subsection{Physical waveform}
\begin{figure}[t]
    \centering
    \includegraphics[width = 0.5\textwidth]{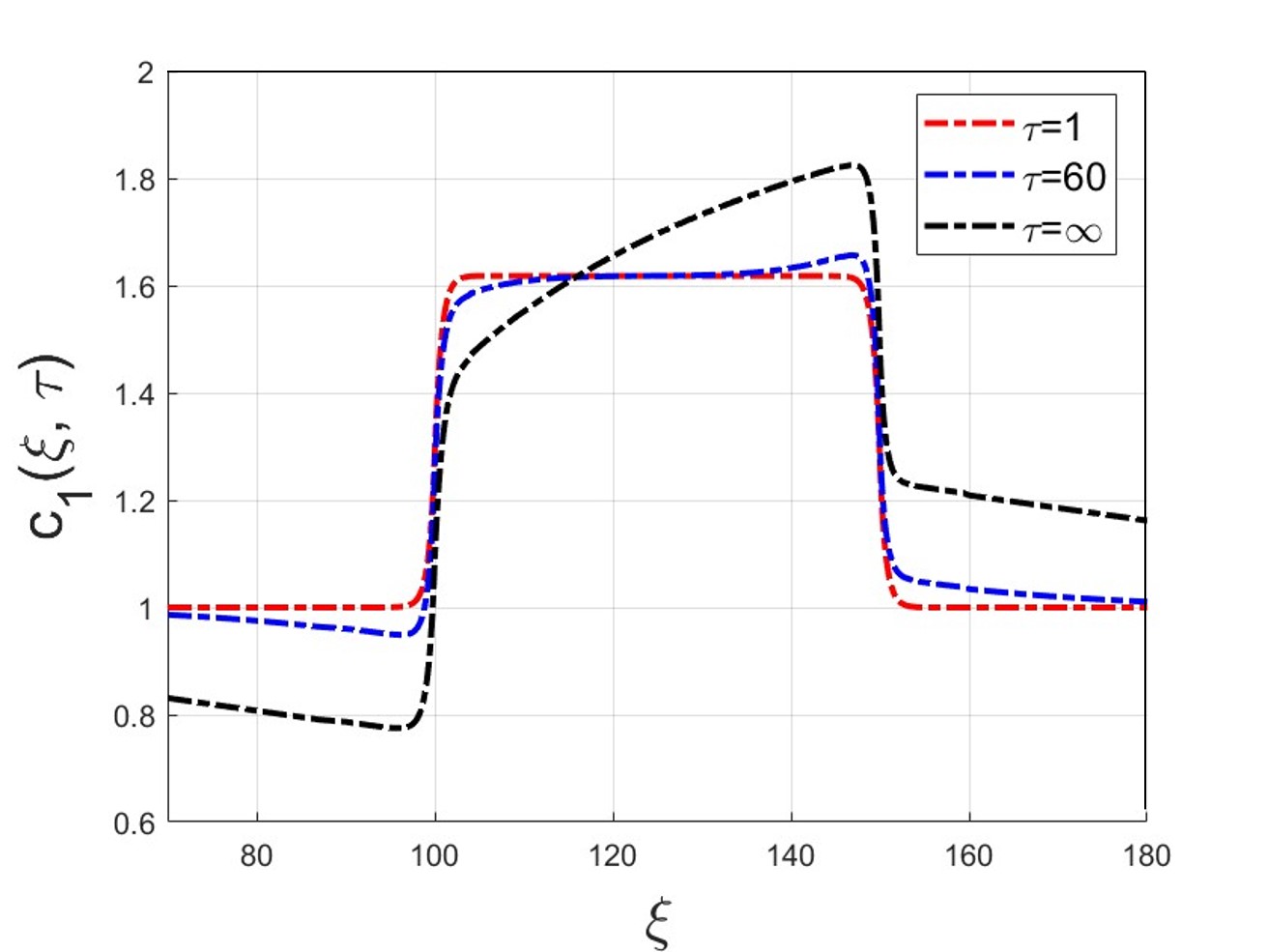}
    \caption{Information molecules concentration profiles in the membrane system recorded at different times in response to a step function of $V_{sig}=5$.}
    \label{Information molecules concentration profiles}
\end{figure}
Fig. \ref{Information molecules concentration profiles} plots the information molecules (cation) concentration profile versus time as a function of position across a cation exchange membrane under an applied voltage. For a small value of this parameter, $\tau = 1$ (red), the profile is flat, with slight polarization. At longer times, $\tau = 60$ (blue), ion migration leads to accumulation on the cathode side and depletion on the anode side. Under steady-state conditions, when $\tau = \infty$ (black), a steep concentration gradient develops due to cation depletion near the anode and significant accumulation near the cathode, exemplifying the fully polarized membrane state under sustained voltage application.

\begin{figure}[t]
    \centering
    \includegraphics[width = 0.5\textwidth]{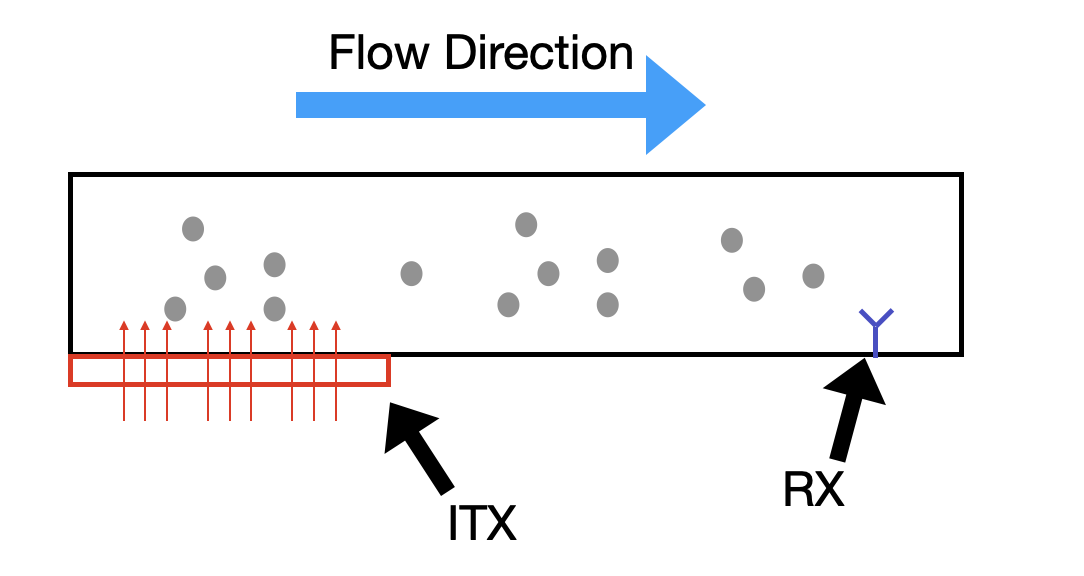}
    \caption{Schematic Representation of 1D MC System: The ITX releases particles into the medium, which propagate towards the receiver (RX) under the influence of flow in the indicated direction $y$.}
    \label{Schematic Representation of 1D Molecular Communication System: The transmitter (TX) releases particles into the medium, which propagate towards the receiver (RX) under the influence of flow in the indicated direction.}
\end{figure}

From communication engineer perspective, we are more interested in the ion flux coming out of the membrane instead of the concentration profile. The ion flux out of the membrane in network model will just be the current between the membrane and the solution. Since the simulation is carried out to model the 1D membrane ion transport, as shown in Fig. \ref{Schematic Representation of 1D Molecular Communication System: The transmitter (TX) releases particles into the medium, which propagate towards the receiver (RX) under the influence of flow in the indicated direction.}, the ion flux produced in each simulation can be consider as the flux from a point ITX source. The problem could be further reduced to a 1D environment.The expected concentration due to  a point ITX for a 1D passive receiver can be expressed as  

\begin{equation}
\bar{C}_{\text{point}}(\tau) = \frac{J(y, \tau)}{\sqrt{4 \pi D \tau}} \exp \left(-\frac{(d - u \tau)^2}{4 D \tau}\right),
\end{equation}
where $y$ is the axis in the flow direction and $u$ is the flow speed \cite{crank1979mathematics}.

In the membrane communication 1D case, the impulse response for a line ITX can be obtained by summing the contributions from all discretized point sources along the line:

\begin{equation} 
\bar{C}_{\text{line}}(\tau) = \sum_{k=0}^{n-1} C_k(\tau).
\end{equation}

Assuming each segment has the same transmission characteristics, the final concentration waveform at the observation point can be approximated as:
\begin{multline}
    \bar{C}_{\text{line}}(\tau_i) \approx \sum_{k=0}^{n-1} \sum_{j=0}^i J_k(\tau_j) \frac{\Delta \tau}{\sqrt{4 \pi D (\tau_i - \tau_j)}} \\
    \times \exp \left(-\frac{(y_{\text{obs}} - y_k - u (\tau_i - \tau_j))^2}{4 D (\tau_i - \tau_j)}\right),
\end{multline}
where \( i \) represents the index for the current observation time \( \tau_i \), indicating the specific time point at which the total concentration \( \bar{C}_{\text{line}}(\tau_i) \) is being evaluated. The $\Delta \tau$ represents the discrete interval between successive time points (time step). The index \( j \) represents past time steps up to \( \tau_i \), referring to the specific time step \( \tau_j \) considered in the summation to calculate contributions to the concentration at time \( \tau_i \). The range \( j = 0 \) to \( i \) ensures that only contributions up to the current observation time are included in the calculation. The term \( u(\tau_i - \tau_j) \) represents the displacement due to flow speed \( u \) over the time interval \( (\tau_i - \tau_j) \), introducing an advection component that shifts the concentration profile along the direction of flow.

\begin{figure}[t]
    \centering
    \includegraphics[width = 0.45\textwidth]{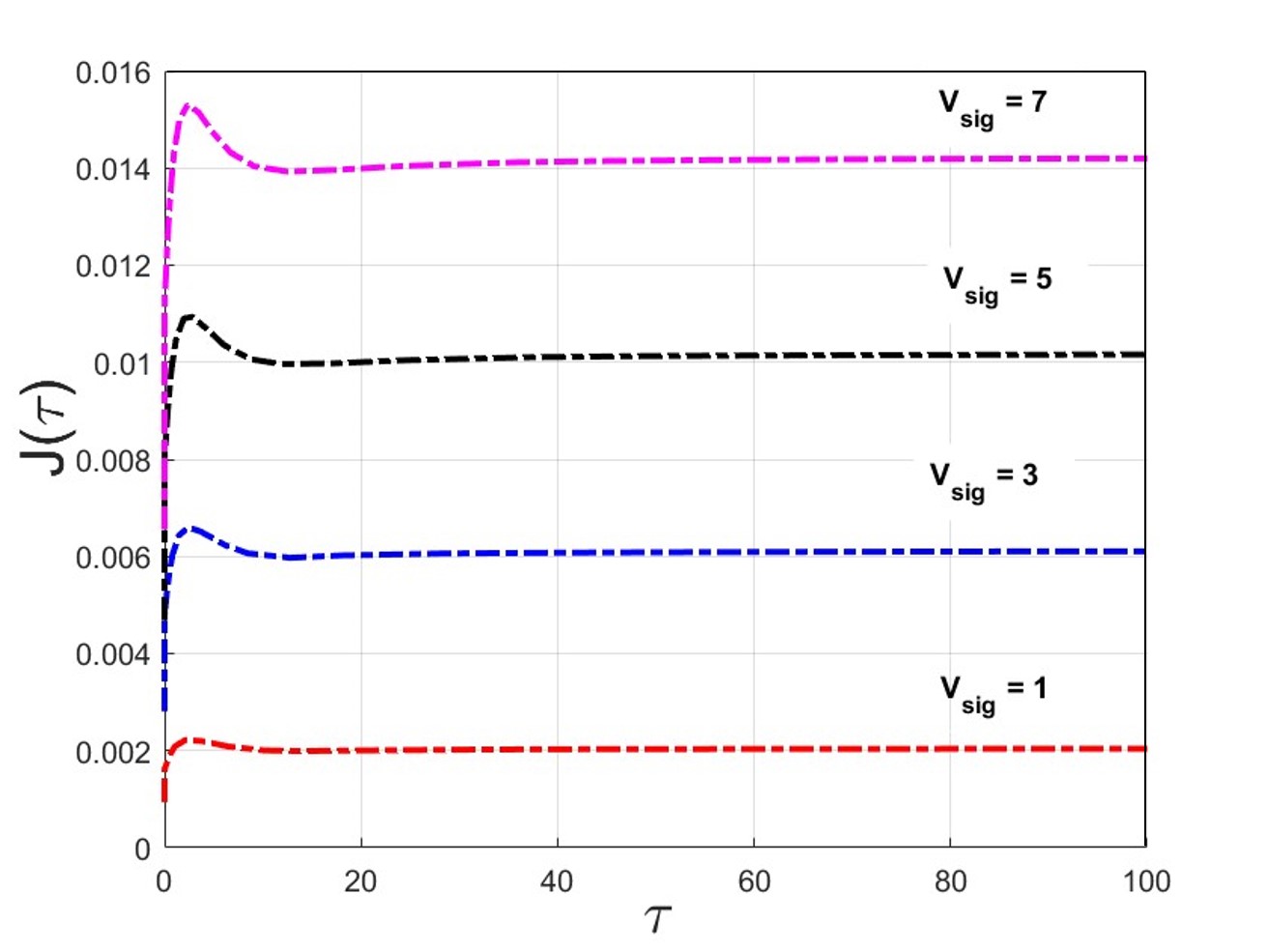}
    \caption{Temporal Variation of Flux \( J(\tau) \) for Different Signal Voltages \( V_{\text{sig}} \): The plot illustrates the flux response over time for varying signal voltages \( V_{\text{sig}} = 1, 3, 5, \) and \( 7 \), showcasing how the flux amplitude changes as the input signal voltage increases.}
    \label{Temporal Variation of Flux.}
\end{figure}

Fig. \ref{Temporal Variation of Flux.} shows the temporal variation of flux \( J(\tau) \) for different input signal voltages \( V_{\text{sig}} \). This figure is essential for understanding how the system responds dynamically under various signal conditions and the time it takes for the flux to reach a steady state. The characteristic time scale \( \tau \) describes how quickly the system transitions from the initial transient phase to a steady state. This time scale depends on several factors, such as the diffusion coefficient \( D_{iM} \), the length scale \( L \), and the applied electric potential. As mentioned earlier, (\ref{characteristic time constant}) aligns well with the flux plot, which is calculated to be around 10. This indicates that the assumptions made earlier are valid for this system.

\begin{figure}[t]
    \centering
    \includegraphics[width = 0.5\textwidth]{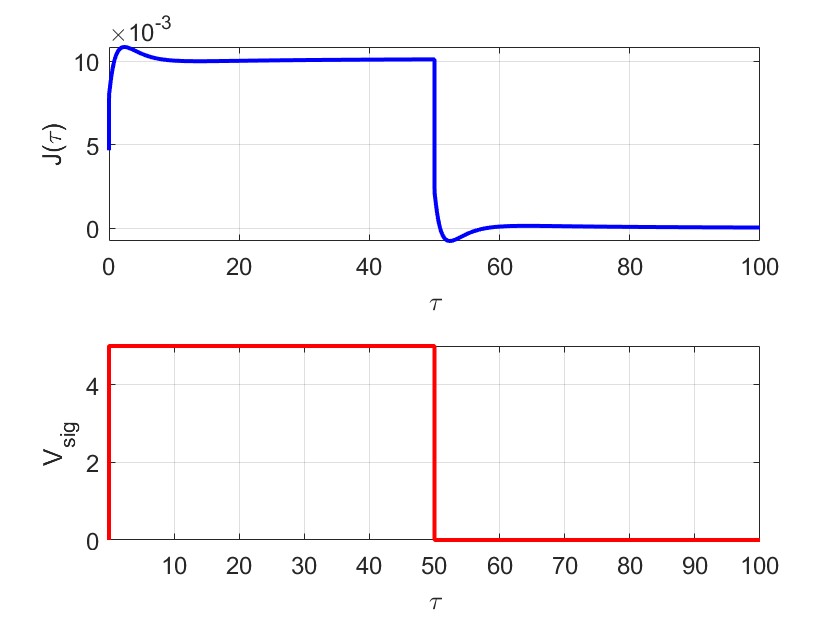}
    \caption{Comparison of Flux Response \( J(\tau) \) and Input Signal \( V_{\text{sig}} \) Over Time. The top plot shows the behavior of the flux \( J(\tau) \) over the time interval \( \tau = 0 \) to \( \tau = 100 \)}
    \label{squarewave}
\end{figure}

Fig. \ref{squarewave} illustrates the response of the flux \( J(\tau) \) to a square wave input \( V_{\text{sig}} \). At the start, when \( V_{\text{sig}} \) transitions from 0 to a positive value, \( J(\tau) \) shows an immediate increase, reflecting the system’s rapid response. This phase is marked by a transient rise as the flux accelerates towards a new steady-state value. Once the input stabilizes at its positive level, \( J(\tau) \) reaches a plateau, maintaining a steady flux output that corresponds to the sustained input signal.

When \( V_{\text{sig}} \) returns to 0, \( J(\tau) \) experiences a sharp decline, entering a negative transient phase as the system adapts to the sudden reduction. This may involve an undershoot where \( J(\tau) \) briefly dips below the baseline before recovering. Finally, the flux gradually returns to its baseline (or zero) level, staying stable until the next positive shift in the square wave.

The shape of \( J(\tau) \) for the square wave input highlights both the transient dynamics and steady-state behavior of the system. The shape and rate of these changes are influenced by diffusion properties, system time constants, and any reactive components present.

\begin{figure}[!ht]
\centering
\includegraphics[width=0.45\textwidth]{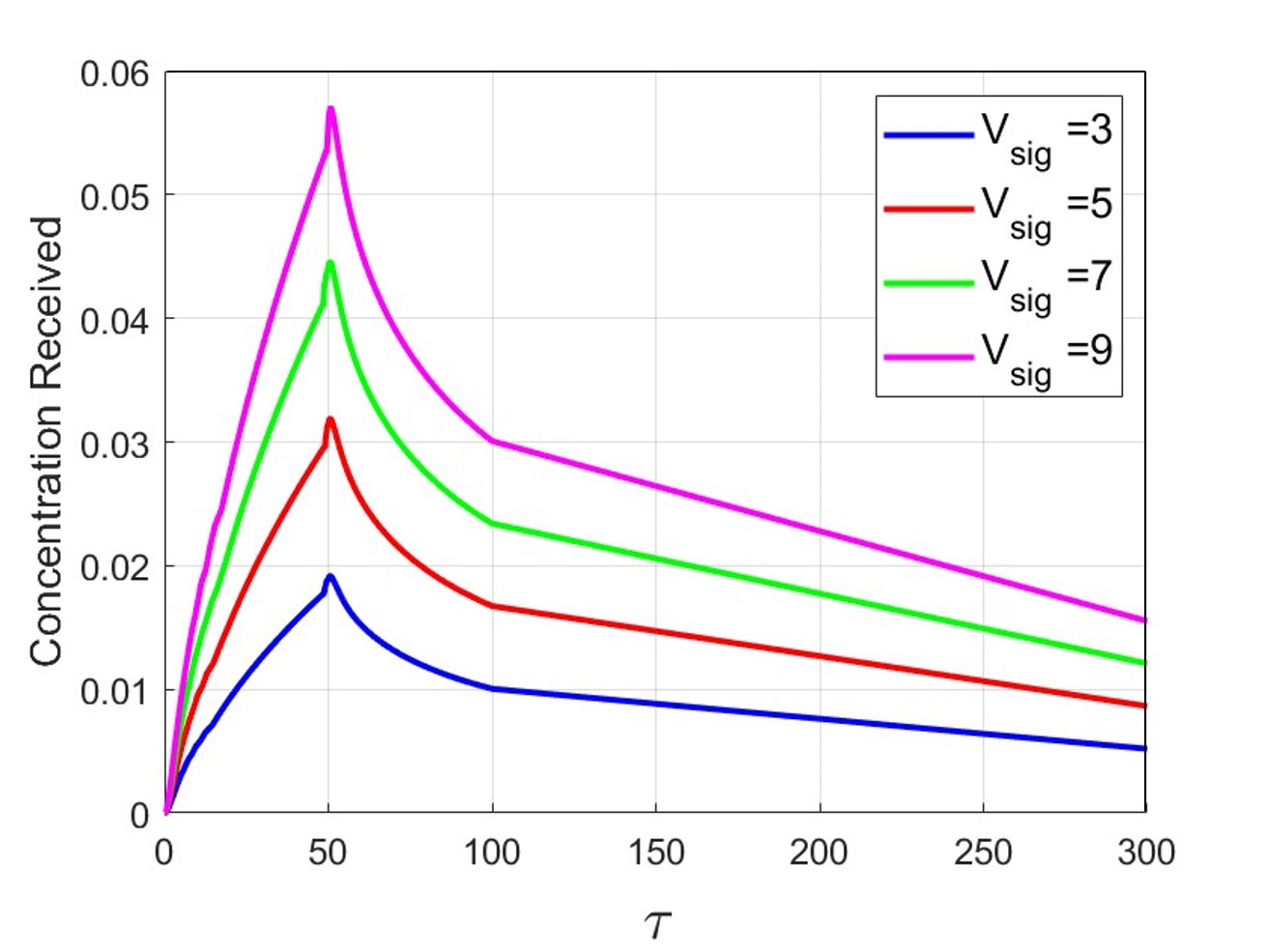}
\caption{Concentration Received at the Observation Point for Different Square Input Signal Voltages \( V_{\text{sig}} \). The plot shows the received concentration profiles over time \( \tau \) for various input signal amplitudes, \( V_{\text{sig}} = 3, 5, 7, \) and \( 9 \).}
\label{fig:concentration_waveform}
\end{figure}

Fig. \ref{fig:concentration_waveform} presents the concentration profiles received at the observation point for different input signal amplitudes \( V_{\text{sig}} \) of 3, 5, 7, and 9. The figure illustrates that as the signal amplitude increases, the peak concentration also rises, indicating a stronger response. 

\subsection{SNR analysis}

\begin{figure}[!ht]
\centering
\includegraphics[width=0.45\textwidth]{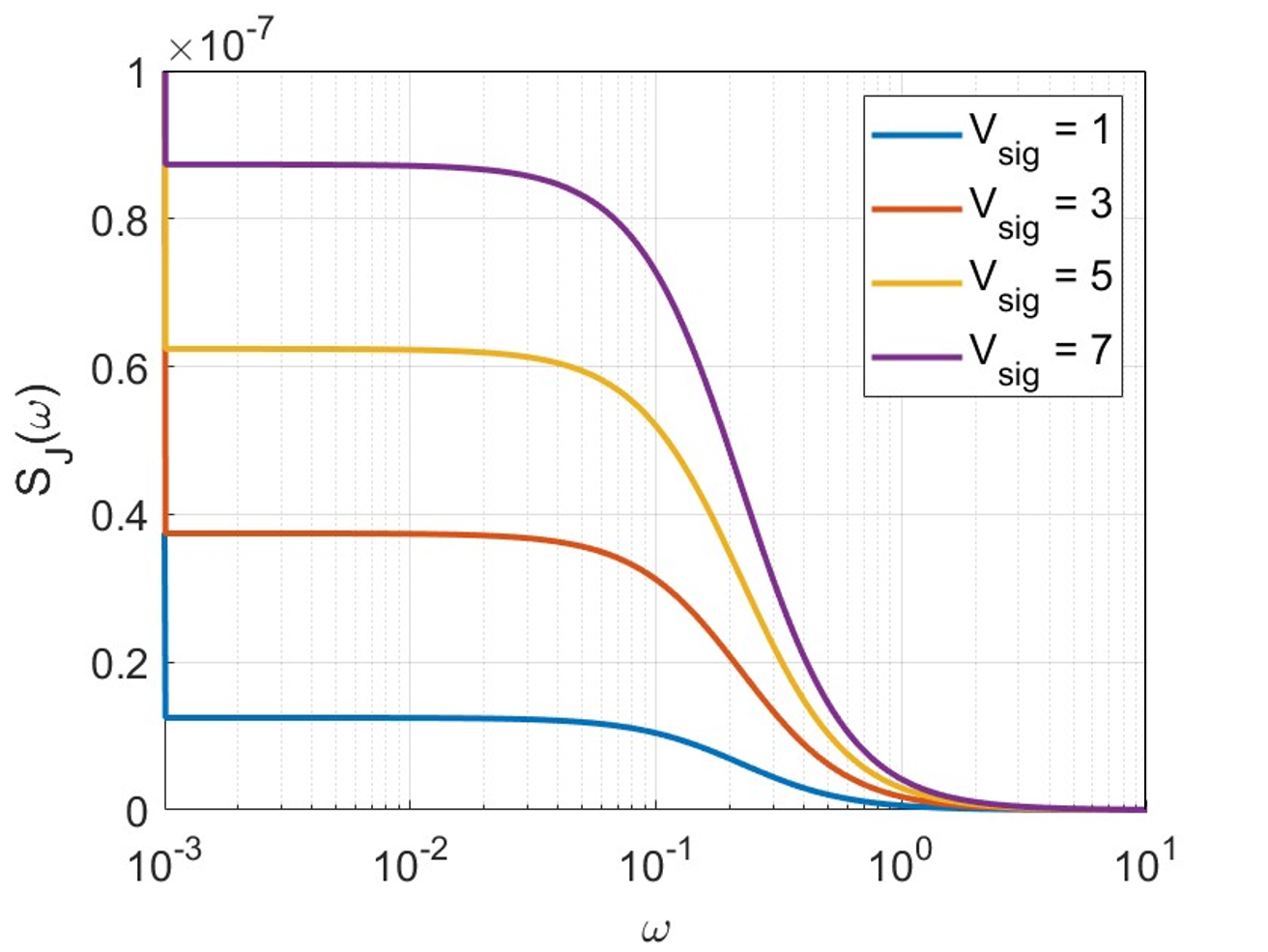}
\caption{Spectral noise density $S_J(\omega)$ for different values of $V_{sig}$. The plot shows how the noise power changes with frequency for varying values of the signal voltage $V_{sig}$ with $D_a = 2\times 10^{-9}$, $c_{bulk} = 100$ and $\epsilon_m =2 $.}
\label{Spectral noise density}
\end{figure}

\begin{figure}[!ht]
\centering
\includegraphics[width=0.45\textwidth]{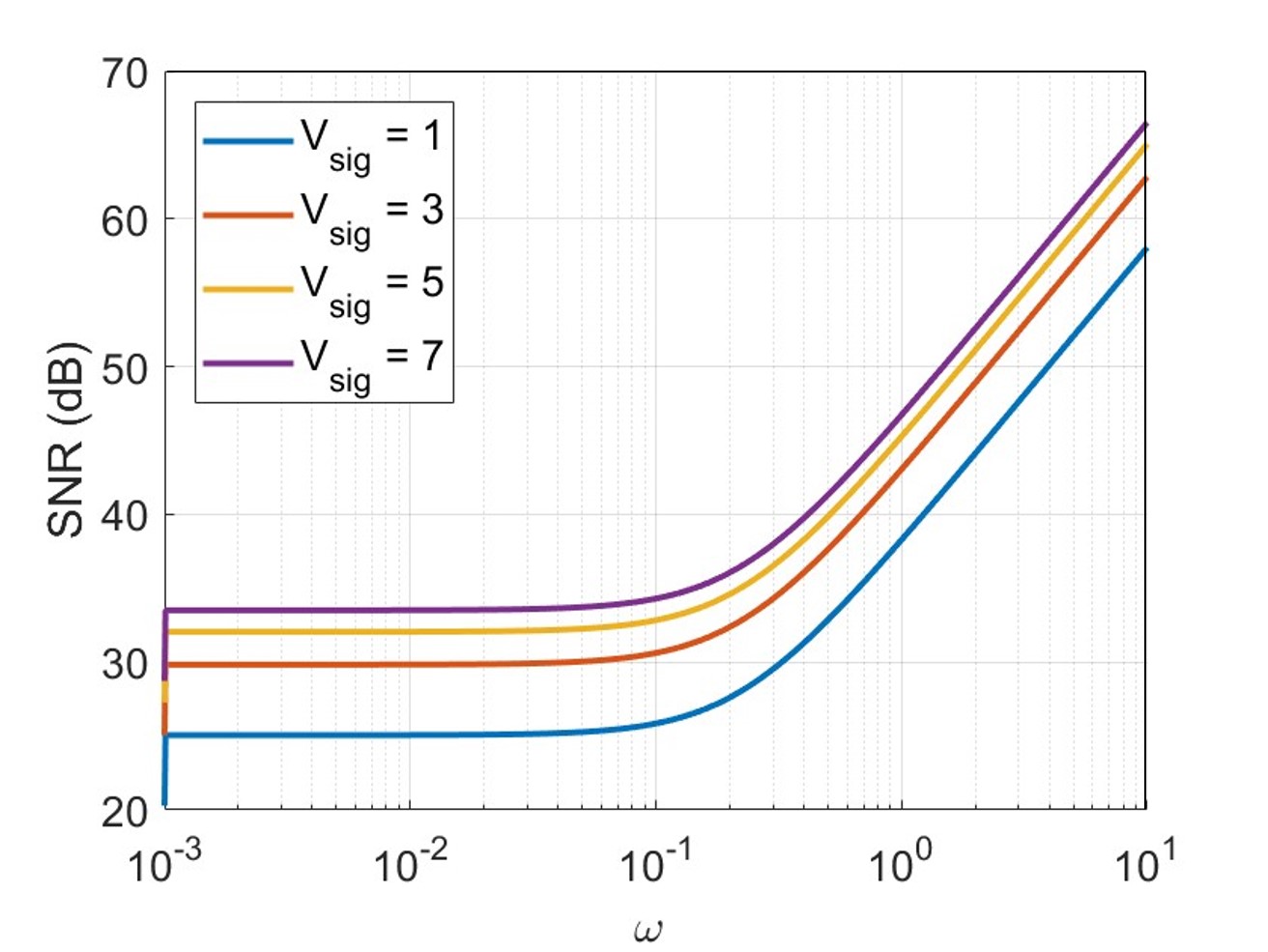}
\caption{Signal-to-Noise Ratio (SNR) as a function of frequency $\omega$ for different values of the signal voltage $V_{sig}$ with $D_a = 2\times 10^{-9}$, $c_{bulk} = 100$ and $\epsilon_m =2 $. The plot demonstrates how SNR changes with increasing frequency, highlighting the impact of varying $V_{sig}$ on system performance.}
\label{SNR}
\end{figure}

\begin{figure}[!ht]
\centering
\includegraphics[width=\linewidth]{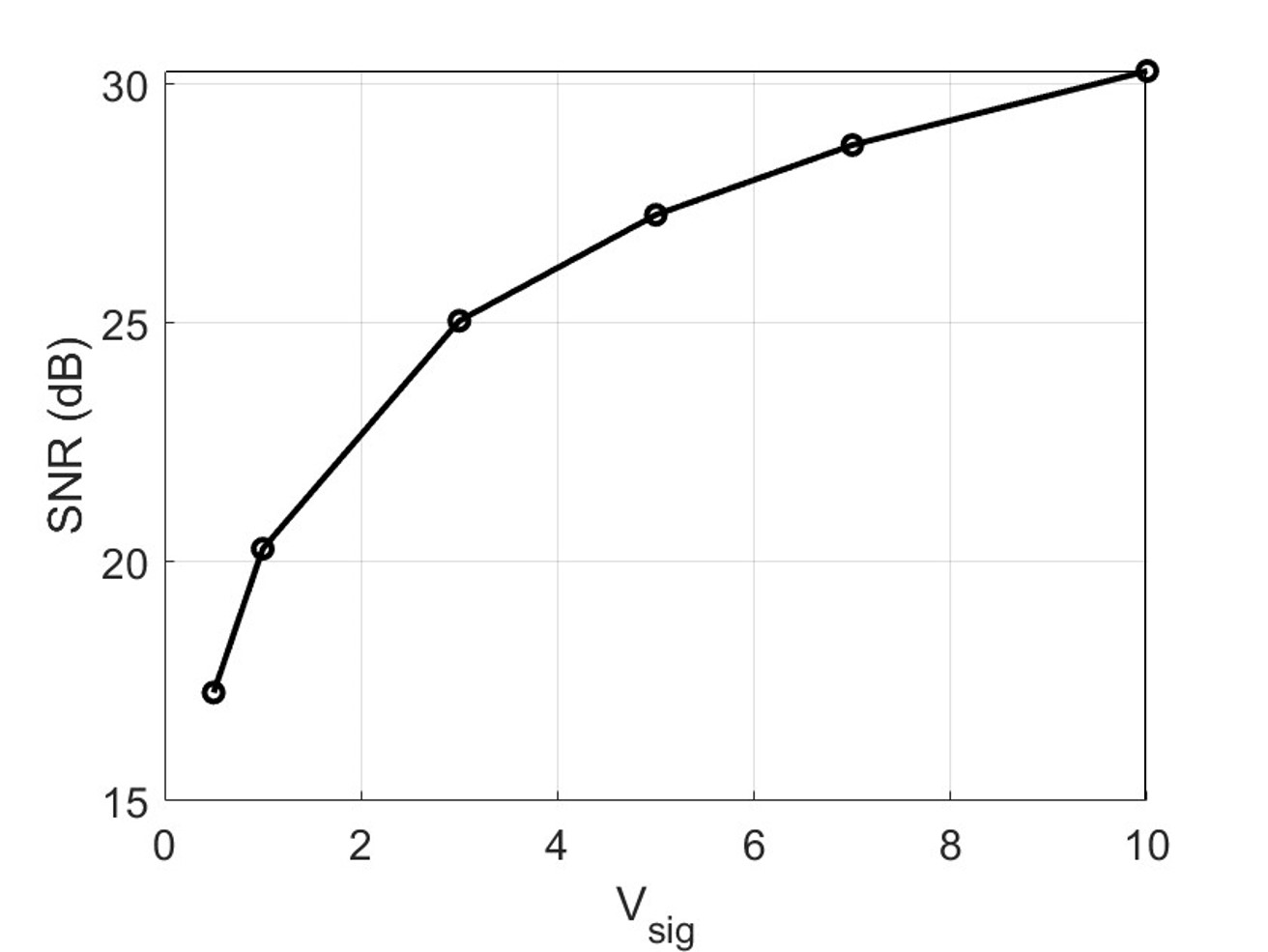}
\caption{Relationship between Signal-to-Noise Ratio (SNR) and Input Signal Amplitude \( V_{\text{sig}} \) at DC with $D_a = 2\times 10^{-9}$, $c_{bulk} = 100$ and $\epsilon_m =2 $. The plot illustrates how the SNR, expressed in decibels (dB), changes as the input signal amplitude \( V_{\text{sig}} \) increases.}
\label{fig:vsig_snr}
\end{figure}

Fig. \ref{Spectral noise density} presents the spectral noise density \( S_J(\omega) \) as a function of dimensionless frequency $\omega$ for input signal amplitudes \( V_{\text{sig}} \) ranging from 1 to 7. One can notice that in all cases, \( S_J(\omega) \) is almost constant in the low-frequency region and decreases abruptly at $\theta$ as frequency increases. Larger values of \( V_{\text{sig}} \) correspond to a greater magnitude of noise power, suggesting that an increase in input signal amplitude contributes to higher noise power at low frequencies.

Fig. \ref{SNR} shows the signal-to-noise ratio (SNR) as a function of dimensionless frequency \( \omega \) for the same set of input signal amplitudes \( V_{\text{sig}} \). The SNR plot demonstrates that higher \( V_{\text{sig}} \) results in improved SNR across the entire frequency range, with the SNR curve rising more significantly at higher frequencies. This suggests that higher input voltages enable the signal's strength to better overcome the noise, especially at higher frequencies. Considering the time constant of the ITX within the operating bandwidth, noise exhibits a flat power spectral density as does the SNR.

Finally, Fig. \ref{fig:vsig_snr} illustrates the relationship between SNR and the input signal amplitude \( V_{\text{sig}} \) when $\omega=0$. The plot indicates a non-linear increase in SNR with rising \( V_{\text{sig}} \). The rate of SNR improvement becomes more gradual at higher \( V_{\text{sig}} \) values, implying diminishing returns as the input signal amplitude continues to increase.

Overall, these findings indicate that while higher \( V_{\text{sig}} \) improves SNR and noise power, the rate of improvement slows at higher amplitudes. This behavior is crucial for optimizing system performance by balancing signal strength and noise characteristics.

\section{Future Direction and Conclusion}
\label{Future Direction and Conclusion}
Future studies should focus on developing strategies to mitigate noise and improve signal integrity in ion exchange membrane-based communication systems. Noise, particularly thermal and shot noise, poses significant challenges that degrade the signal-to-noise ratio (SNR) and impact communication reliability. Exploring innovative modeling approaches and advanced signal processing algorithms may help minimize these noise sources and enhance overall system performance. Additionally, addressing low-frequency noise, such as $1/f$ noise, is essential for realistic ITX design when the membrane is no longer homogeneous. An analytical study of flicker noise in synthetic membranes remains largely unexplored and could offer valuable insights into noise modeling.

Improving the modeling on selectivity is another key area for future research. Robust signal transmission depends on accurately simulating ion transport, while selectivity modeling ensures that desired ions contribute effectively to the signal. Current models of ion exchange membranes typically focus on differentiating molecules based on valency, which may not adequately capture complex real-world interactions. Future research could explore more detailed simulations that incorporate dynamic changes in membrane behavior and interactions with various ion types. Investigating how these properties affect signal transmission and implementing adaptable models that respond to system condition changes could enable more effective and versatile membrane-based communication frameworks.

Environmental sensitivity presents additional challenges in modeling ion exchange membranes for communication applications. In our current model, the channel adjacent to the membrane is designed with a controlled concentration of information molecules to minimize the impact of free diffusion contributing to the signal. However, modeling must account for varying external conditions, such as different side channel concentrations, temperature and chemical composition changes, which can affect performance. Future modeling efforts should incorporate these environmental factors and simulate their effects on ion transport and communication efficiency. Developing adaptive simulation frameworks that dynamically adjust to reflect real-time environmental changes could further enhance the accuracy and reliability of membrane-based communication system models.

In conclusion, this paper proposes and evaluates physical MC ITXs based on synthetic ion exchange membranes and analyzes their performance using network simulation methods. The results indicate that ion exchange membranes are feasible for MC, with particular attention to SNR performance and waveform analysis. Although the findings are promising, challenges such as noise mitigation and enhancing membrane selectivity need to be addressed to achieve practical implementation. Future research should focus on advanced modeling and adaptive frameworks to further enhance system reliability and performance under varying environmental conditions.

\bibliographystyle{IEEEtran}
\bibliography{references}
\vfill
\end{document}